\documentclass[sigconf]{acmart}

\usepackage{booktabs} 
\usepackage{amssymb}
\usepackage{graphicx}
\usepackage{algorithm}
\usepackage{algorithmic}
\usepackage{multirow}
\usepackage{makecell}
\usepackage{url}

\acmConference[Social Web in Emergency and Disaster Management]{ACM WSDM conference}
\acmYear{2018}

\begin{document}
\title{A Pipeline for Post-Crisis Twitter Data Acquisition}

\author{Mayank Kejriwal}
\affiliation{%
  \institution{Information Sciences Institute}
  \streetaddress{USC Viterbi School of Engineering}
  \city{Marina Del Rey} 
  \state{California} 
  \postcode{90292}
}
\email{kejriwal@isi.edu}

\author{Yao Gu}
\affiliation{%
  \institution{Department of Computer Science}
  \streetaddress{University of Southern California}
  \city{Los Angeles} 
  \state{California} 
  \postcode{90089}
}
\email{yaogu@usc.edu}






\renewcommand{\shortauthors}{Kejriwal and Gu}

\begin{abstract}
Due to instant availability of data on social media platforms like Twitter, and advances in machine learning and data management technology, \emph{real-time crisis informatics} has emerged as a prolific research area in the last decade. Although several benchmarks are now available, especially on portals like CrisisLex, an important, practical problem that has not been addressed thus far is the rapid acquisition and benchmarking of data from free, publicly available streams like the Twitter API. In this paper, we present ongoing work on a pipeline for facilitating immediate post-crisis data collection, curation and relevance filtering from the Twitter API. The pipeline is \emph{minimally supervised}, alleviating the need for feature engineering by including a judicious mix of data preprocessing and fast text embeddings, along with an active learning framework. We illustrate the utility of the pipeline by describing a recent case study wherein it was used to collect and analyze millions of tweets in the immediate aftermath of the Las Vegas shootings.   
\end{abstract}

%
%



\maketitle

\section{Introduction}

In recent years, crisis informatics has emerged as a field unto its own due to a growing recognition that technology, especially intelligent systems, can be used to better mobilize resources and provide valuable, rapid insights to field operators and analysts in the aftermath of a crisis \cite{policyforum}. Crises include not just natural disasters, but also human-mediated disasters caused by terrorism or shootings. 

An important reason that technology can help in such situations is the availability of data in real-time from social media platforms like Twitter. In recent years, several efforts, such as CrisisLex \cite{olteanu2014crisislex}, have confirmed that useful crisis data is available on Twitter. While these benchmark collections can be used for detailed posthoc analyses, as well as pre-training machine learning systems, an open research problem in the crisis informatics literature is the rapid acquisition, preprocessing and, potentially, analysis and visualization, of a crisis-specific dataset in the aftermath of a \emph{new} crisis. We assume in this paper that the `raw stream' from which this data has to be acquired is the Twitter public API, which is available for free and can be used to collect data up to a limit set by Twitter. In relative terms, this limit is not large, but due to the massive size and scope of Twitter at any given time (over 500 million tweets are estimated to be globally streamed per day), as well as the querying capabilities offered by the API, it is still a valuable resource that can be availed of, free of charge. 

Unfortunately, although the stream can be used to acquire a high-recall dataset by optimistically specifying keywords and hashtags in the query, a dataset acquired in this way is not high-precision. Namely, there are many tweets in the acquired dataset that are \emph{irrelevant} and only contribute noise to the overall corpus. As a running example, consider the crisis of the Las Vegas shootings, which occurred (on the night of Oct. 1) in the vicinity of the Mandalay Bay resort where the gunman had a room. To achieve sufficient recall, one would have to use keywords like `las vegas' and `mandalay bay' in the Twitter API, but either keyword can (and does) lead to tweets that have nothing to do with the shooting. The problem is further compounded when one considers data acquisition over time. In the immediate aftermath of the Las Vegas shootings, for example, it is much more likely for a `las vegas' tweet to be relevant to the shooting than several days later. Thus, precision declines over time, making the acquisition of a high-quality, relevant dataset over a non-trivial time period even more challenging than usual. If such a dataset can be acquired and analyzed in near real-time, it would significantly aid practitioners and field operators looking to obtain \emph{situational awareness} into the crisis as it is unfolding \cite{vieweg2010}. 
\begin{figure}
\centering
\includegraphics[height=2.0in, width=3.4in]{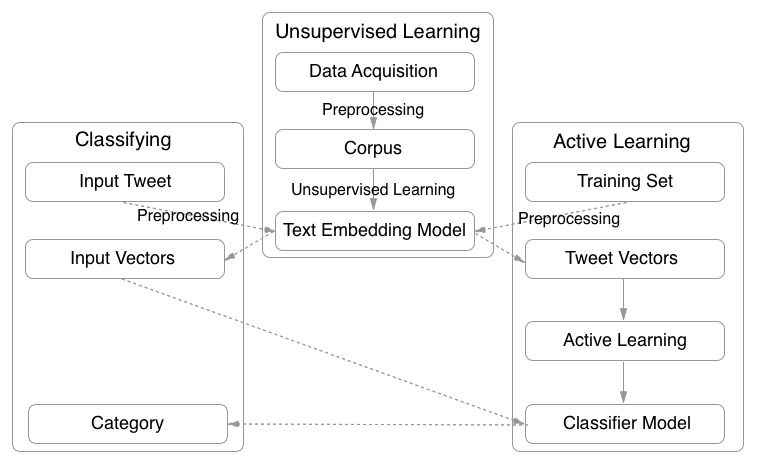}
\caption{A workflow-level illustration of the data acquisition pipeline.}\label{fig:approach}
\end{figure}

In this paper, we present ongoing work on an end-to-end data acquisition pipeline (Figure \ref{fig:approach}) that can be used to collect a crisis-specific dataset from the Twitter public API using minimal human supervision. The pipeline includes key steps such as data preprocessing and filtering, but does not require a user to engineer features or label thousands of tweets before delivering meaningful results. Instead, as illustrated in Figure \ref{fig:approach}, we leverage a judicious combination of unsupervised text embeddings and active learning to acquire the dataset with minimal human engineering. Labeling effort is restricted to a few tens of samples, in addition to interactively specifying keywords and hashtags to obtain an initial data corpus from Twitter when the crisis first strikes.

{\bf Contributions.} The two main contributions of this work are as follows. First, we present ongoing work on a simple and scalable end-to-end pipeline that ingests data from the Twitter streaming API and uses a combination of unsupervised text embeddings and limited-label active learning to construct a crisis-specific training set. Second, we use a case study around the Las Vegas shooting massacre that occurred on the night of Oct. 1 to illustrate the promise of the approach. Compared to a baseline control, active learning is found to converge faster in the text embedding space. Text embeddings are found to yield intuitive results, illustrating the combined robustness of data collection and preprocessing. 

{\bf Structure of the paper.} Section \ref{relatedwork} covers relevant related work, while Section \ref{approach} describes the key components in Figure \ref{fig:approach}. Section \ref{experiments} presents a brief set of preliminary empirical results using a case study, namely the Las Vegas shootings. Section \ref{conclusion} presents promising avenues for future work and concludes the paper.


\section{Related Work}\label{relatedwork}

Crisis informatics is emerging as an important field for both data scientists and policy analysts. A good introduction to the field was provided in a recent Science policy forum article \cite{policyforum}. The field draws on interdisciplinary strands of research, especially with respect to collecting, processing and analyzing real-world data. We cover some relevant work below.

\subsection{Social Media and Crisis Informatics}
Social media platforms like Twitter have emerged as important channels (`social sensors' \cite{earthquake1}) for situation awareness in socially consequential domains like crisis informatics. While the initial primary focus was on earthquakes \cite{earthquake2}, \cite{earthquake3}, the focus has diversified in recent years to disasters as diverse as floods, fire, and hurricanes \cite{floods}, \cite{vieweg2010}. We note that Twitter is by far the most monitored social media platform during crises \cite{simon2015socializing} due to the availability of the published data and its real-time nature. Increasingly sophisticated approaches have been presented for data collection, including dynamic lexicons \cite{olteanu2014crisislex}. For free use of streaming Twitter data, the public API is the single point of query access. Since data collection in this paper focuses on a single disaster, we assume that a user has access to some keywords that she would use to input initial queries to the API. An advantage of our approach is that, because we aim to improve precision subsequently, the user can afford to be liberal and optimistic in her choice of query keywords and hashtags.

\subsection{Data Preprocessing}
Social media content and text is generally heterogeneous, with unusual spellings and language models. Systems that have been found to work well for Twitter have employed a variety of preprocessing steps \cite{pre1}, \cite{pre2}, \cite{pre3}. The system in this paper also employs some preprocessing steps.

\subsection{Data Filtering and Curation}
An important initial step when dealing with heterogeneous information sources is to separate \textit{relevant} (i.e. crisis-related) and \textit{irrelevant} documents\cite{olteanu2015expect}. This allows the filtering of documents that may have used a \textit{crisis-related} term or hashtag, but does not contain information that is relevant to a \emph{particular} crisis event. Important filtering methods have been covered by a range of papers, such as \cite{dualcnn}, \cite{semcnn} and \cite{prashant}. Importantly, the filtering in this paper assumes little domain knowledge and minimal labeling effort from the user. A user is also not required to program a machine learning system or devise inventive features. This enables benchmarks to be quickly collected in a matter of hours rather than days or weeks, and could potentially be leveraged by efforts like CrisisLex to significantly expand their current benchmark collection.

\subsection{Data Analysis}
Although the primary focus of this paper is on data acquisition, preprocessing and relevance filtering, the ultimate goal of acquiring such a dataset is to conduct analysis. Many such analyses require the underlying dataset to be composed of relevant documents, though some methods are more robust to noise than others. Analysis tasks include (1) \emph{event detection} (both extraction and co-reference) \cite{rees}, \cite{ritter}, \cite{grishman}, for which a variety of machine learning techniques have been proposed \cite{recent1}, \cite{recent2}, \cite{recent3}, and that was surveyed in \cite{edsurvey}; (2) \emph{data classification}, which concerns identifying the \emph{type} of information expressed in a document (e.g., \textit{donations and volunteering, infrastructure and utilities, affected individuals}) \cite{olteanu2015expect}, since such knowledge can be used by responders to provide actionable information that is generally missing from general event categories. Deep learning methods have recently shown a lot of promise in automatically identifying such information \cite{dualcnn}, \cite{semcnn}; (3) \emph{named entity recognition}, especially for entity-centric querying (e.g., to answer questions such as \emph{what are the sentiments associated with the UN in Ghana?}) to acquire finer-grained situational awareness \cite{entity1}, \cite{entity2}, \cite{entity3}; (4) visualization, which is an important part of any human-centric system that is attempting to make sense of a large amount of information. Several good crisis informatics platforms that provide visualizations include \cite{ushahidi}, \cite{twitris}, \cite{twitcident}, \cite{aidr}, \cite{crisistracker}, \cite{tweettracker}, \cite{choi2015real}, \cite{thom2015can}.

We present preliminary results on visualization, which is the most important component of a forward-facing crisis informatics system that necessarily involves humans in the loop. The visualization is generated in an unsupervised fashion, has a simple, interactive component that summarizes the corpus using \emph{hashtags} as visual units, and is freshly rendered for every new disaster. In other words, it does not require pre-customization or extensive set-up. 

\section{Approach}\label{approach}

The overall approach is illustrated in Figure \ref{fig:approach} and is quite simple. We assume that the crisis has just struck, and the user has obtained some clues from an external source (e.g., a local breaking news, or privileged first responder information). As a first step, the user has to specify some inputs so that the Twitter API can start retrieving and storing a corpus in real-time from the stream. Details on this search service may be found at the following\footnote{\url{https://developer.twitter.com/en/docs/tweets/search/overview}} link; details on the API\footnote{\url{https://developer.twitter.com/en/docs/tweets/sample-realtime/overview}} are also available. 
We serialize the real-time streaming inputs by storing the tweets in the raw JSON format in which they stream, on disk, while isolating the text component and storing it in an external text file for labeling and text embedding. The former setting (storing the raw tweets) is optional in low-resource settings, where limited computational power or storage is available. Next, we preprocess each tweet by converting all characters in the tweet to lowercase, and stripping out URLs, user mentions and hashtags. The preprocessing can also be done on a per-tweet basis in real-time. To take a real-world example of tweet preprocessing, the tweet `RT @TheLeadCNN: Remembering Rocio Guillen Rocha, from Anaheim, California. \#LasVegasLost \url{https://t.co/QuvXa6WvlE https://t.co/Og5HpQqUCC}', after preprocessing, would become ` remembering rocio guillen rocha from anaheim california'.

Currently, the next few steps take place in batch mode, but online options for many of these algorithms are available in the literature and are in the process of being integrated into the pipeline. We assume that a preprocessed corpus, serialized as text files, is available on disk. We execute the unsupervised fastText word embedding package on this corpus to obtain a semantic embedding model for the corpus vocabulary \cite{fasttext}. FastText has some notable advantages which make it suited for Twitter data. First, as the name suggests, the model is very fast, and is well-suited for millions of tweets and documents. Second, the model offers convenient modeling and querying command line APIS. Finally, and most importantly, the model is robust, and can be used to deduce vectors for words that were not seen during training. Since Twitter language has high variance and can be quite irregular, we believe that this facility is paramount to inferring good text embeddings.  

\begin{figure}
\centering
\includegraphics[height=2.5in, width=3.1in]{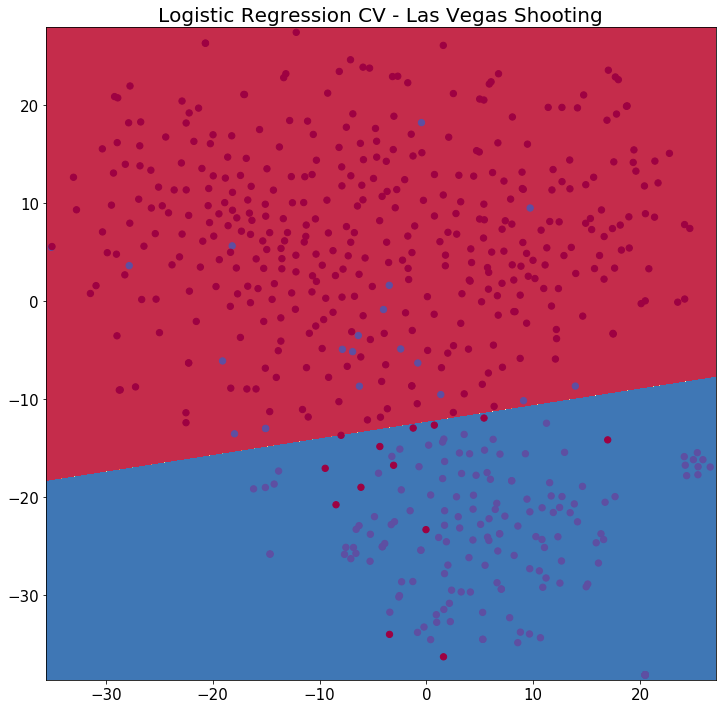}
\caption{A regression plot from our Las Vegas case study experiments. While we illustrate the true labels (i.e. points are colored blue or red) in the figure, the active learning method would not have access to this information and would pick the points closest to the line as the `uncertain' data for the next iteration.}\label{fig:reg}
\end{figure}

Thus far, all steps were completely unsupervised. In order to acquire the benchmark and jumpstart the active learning process, we assume a set of a high-precision heuristics (e.g., a hashtag like `lasvegasmassacre') to sample and manually label a small set (say, 50) of tweets as positive, and another small set of tweets as negative, with positive indicating that the tweet is related in some way to the specific crisis. Using this small `initial pool', we train a machine learning classifier using the text embeddings as features, and bypassing the feature engineering process. We now assume the following active learning framework. First, let us assume that the user labeling budget is X i.e. the user wants to label X more data points. We also assume a hyperparameter $p$, which is the number of samples that the user will label in each active learning iteration; hence, by definition, there will be $X/p$ iterations. In the first iteration, we apply the classifier to the unlabeled pool of data, and select the data that the classifier determines is most uncertain (Figure \ref{fig:reg}). The empirical advantages of using active learning in this way for benchmark construction, as opposed to a baseline control that randomly samples from the unlabeled pool in each iteration, will be illustrated in Section \ref{experiments}.    

\section{Preliminary Experiments: Case Study on Las Vegas Shootings}\label{experiments}

The Las Vegas shooting incident was a major recent tragedy in the city of Las Vegas\footnote{We condense the description provided by the news report detailed in \url{https://www.cbsnews.com/news/more-details-revealed-about-las-vegas-shooters-arsenal-of-weapons/}}. On the night of October 1, 2017, a gunman (64-year-old Stephen Paddock of Mesquite, Nevada) fired more than 1,100 rounds (from his suite on the 32nd floor of the nearby Mandalay Bay hotel) on a crowd of over 22,000 concertgoers at the Route 91 Harvest music festival on the Las Vegas Strip in Nevada, leaving 58 people dead and 546 injured. About an hour after the attack, he was found dead in his room from a self-inflicted gunshot wound. At the time of writing, the motive is unknown, although it has been discovered that he possessed an arsenal of weapons.

According to Wikipedia\footnote{\url{https://en.wikipedia.org/wiki/2017_Las_Vegas_shooting}}, the incident is the deadliest mass shooting committed by an individual in the United States. The shooting reignited the debate about gun laws in the U.S., with attention focused on bump fire stocks, which Paddock used to allow his semi-automatic rifles to fire at a rate similar to that of a fully automatic weapon.

We chose this case study for our analysis because the prototypical pipeline described herein was developed shortly before this shooting took place, and we deployed it shortly after to test its capabilities informally. The tragedy is also recent, and `unusual' and unpredictable in contrast to more weather-related disasters. Towards the end of the paper, we also briefly describe results from another case study concerning hurricanes.    

\subsection{Initial Corpus}

For the preliminary case study experiment, we started collecting data in real time from the Twitter stream on Oct. 3 (i.e. on the second date following the Las Vegas shooting) and terminated data collection on Oct. 9. We used a small set of manually specified keyword phrases such as `las vegas shooting'. We were able to make about 450 requests in every 15 minute window, and obtained roughly 1000 tweets per 15 minute window. The total number of tweets collected over the temporal span of Oct.3-9 is about 1.35 million. 

For purposes of baselining, as well as a second case study, we similarly collected (about 237,480) tweets describing \emph{hurricanes}, including hurricane disasters that various places in the US (e.g., Houston) were still recovering from, as well as hurricanes that had just struck in other places in the world at that time. We also randomly\footnote{We offered the facility offered by the Twitter API, but it is unknown if this facility yields truly random data.} sampled 1.014 million tweets from the API in the date range of Oct. 7-9 to provide a more unbiased negative data for both case studies.

The full corpus (Las Vegas+Hurricane+Random) contains more than 2.6 million tweets and was used to train the fastText embedding, the parameter settings of which we subsequently describe. 





\subsection{Evaluation Dataset Details}

For evaluation purposes, we acquired a manually labeled ground truth of 600 tweets (200 from each of the three corpora previously described) using one of three class labels (\emph{Las Vegas Shooting}, \emph{Hurricane} and \emph{Non-Disaster}). For the Last Vegas experiment, \emph{Hurricane} and \emph{Non-Disaster} were both treated as negatively labeled data, and similarly for the \emph{Hurricane} experiment that we describe at the end of this work. 
The evaluation set is `featurized' by using the preprocessing steps described earlier, and querying for the sentence vector using the unsupervised fastText embedding model trained over the full corpus (of about 2.6 million tweets).

\subsection{Experimental Protocol}

For the fastText text embedding model, we use default parameters for the preliminary experiment and set dimensionality to 100. We use the evaluation set for a limited experiment, wherein we test the effectiveness of the proposed active learning against the non-active learning baseline. Using Logistic Regression for supervised classification, we set up both the active learning (AL) and the baseline control as follows. First, we do stratified sampling of 60\% of the evaluation set, with the positive class comprising Las Vegas shooting-related tweets, and the negative class comprising a union of the other two classes. We use this 60\% for training, and the other 40\% for testing. We \emph{further} split the training set into a 10\%-50\% partition, with the initial 10\% (called the \emph{initial pool}) used for training the \emph{base classifier} (both for AL and the baseline) and the other 50\% used as the \emph{labeling pool}. For the AL experiment, in each iteration, the 5 most \emph{uncertain} data points (according to the probabilities output by the \emph{current} AL classifier) are sampled and merged into the current labeled pool, followed by re-training the classifier. For the baseline control, the protocol is exactly the same, except we \emph{randomly} sample the 5 points from the labeling pool, rather than consider the uncertainty of points when querying.

Using precision and recall for our metrics, we plot, at every iteration, the performance of the current AL and baseline classifiers on the other 40\% (the test set, which is not used by either classifier at any stage for learning). Because we followed this protocol, the results of the baseline and the AL classifier will coincide both at the beginning and at the end. In practice, both the size of the training set, as well as the relative scale of base training and pooling, can both be tuned during evaluation, and we are currently conducting more experiments to analyze such tradeoffs.   

\subsection{Active Learning Results}




\begin{figure}
\centering
\includegraphics[height=2.5in, width=3.5in]{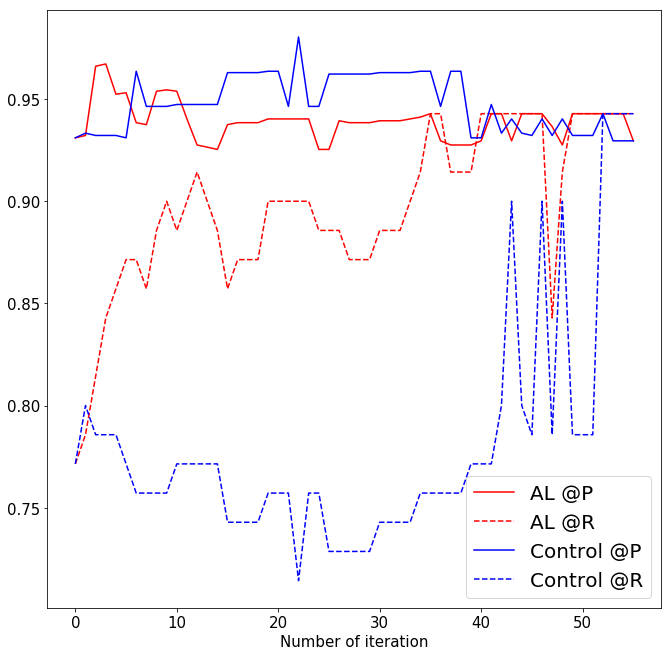}
\caption{Preliminary active learning results for the Las Vegas shooting case study.}\label{fig:al}
\end{figure}
Figure \ref{fig:al} illustrates the results of active learning. In the figure, we use red lines for the active learning and blue lines for the baseline. The solid lines plot precision vs. iterations and dashed lines plot recall vs. iterations. 

Even though it is preliminary, the result is promising because the recall converges much faster than the control group (reaching a stable value just after 10 iterations while the green line is low). Precision continues to be maintained at stable levels as well. This implies that we can use active learning to rapidly acquire a diverse set of data without necessarily sacrificing precision or requiring large-scale training set construction. 

A second case study involving hurricanes, using an identical experimental protocol, illustrated very similar trends (figure not shown herein). Once again, recall was found to significantly improve without eroding precision, and the convergence was much faster for the active learning method.

\subsection{Interactive Hashtag Visualization and Exploration}
\begin{figure*}
\centering
\includegraphics[height=4.3in, width=5.7in]{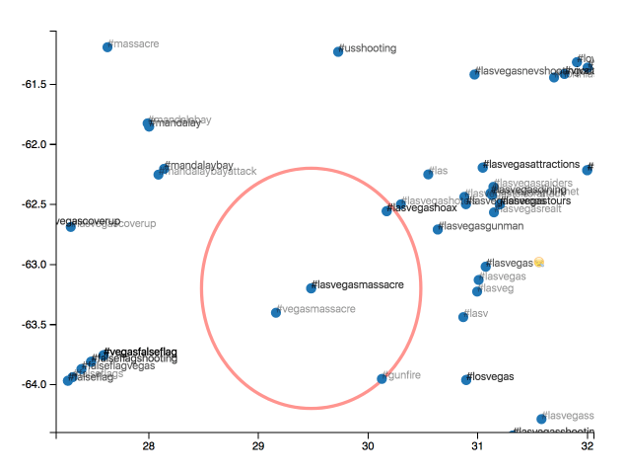}
\caption{Interactively visualizing hashtags using unsupervised text embeddings and t-SNE dimensionality reduction.}\label{fig:hashviz}
\end{figure*}
A concurrent demo submission describes the visualization system in more detail \cite{demo}. The demo uses t-Distributed Stochastic Neighbor Embedding \cite{tsne} for dimensionality reduction in order to visualize high-dimensional tweet vectors and hashtag vectors in 2D space. Similar hashtags will aggregate (implicitly) into clusters, with relative distance in 2D space providing a rough estimate of semantic relatedness. More details are provided in the demo submission \cite{demo}. The example around the Las Vegas massacre shown in Figure \ref{fig:hashviz} shows that the space captures semantic relatedness quite intuitively. Users can interact with the plot using a combination of zooming, scrolling and querying.

\section{Future Work and Conclusion}\label{conclusion}

This paper presented a pipeline for rapid acquisition of a crisis-specific dataset in the immediate aftermath of a crisis. The pipeline uses a small number of interactively labeled samples in an active learning framework, coupled with unsupervised fastText text embeddings, to obtain a relevant corpus without extensive labeling or feature engineering effort. The prototype is being actively developed. Although the results described in Section \ref{experiments} are promising, they are also preliminary. We are looking to validate the pipeline further by considering other case studies, large-scale benchmarking studies and user studies.

\bibliographystyle{ACM-Reference-Format}
\bibliography{sample-bibliography} 

\end{document}